\documentclass{ws-p10x7}
\begin{document}
\title{Hadronic centrality dependence in nuclear collisions}
\author{ Sonja Kabana }
\address{Laboratory for High Energy Physics, University of Bern,
    Sidlerstrasse 5, 3012 Bern, Switzerland, E-mail: sonja.kabana@cern.ch}
\twocolumn[\maketitle\abstract{
\noindent
The kaon number density in nucleus+nucleus and p+p reactions is investigated
for the first time
as a function of the initial energy density $\epsilon$
 and is found to exhibit
 a  discontinuity around $\epsilon$=1.3 GeV/fm$^3$.
This suggests a higher degree of chemical equilibrium for $\epsilon >$
 1.3 GeV/fm$^3$.
It can also be interpreted as reflection of the same
discontinuity, appearing in the chemical 
freeze out temperature (T) as a function of
$\epsilon$.
The  $N^{\alpha \sim 1}$ dependence of (u,d,s) hadrons, with N the
number of participating nucleons,
also indicates a high degree of chemical equilibrium and T saturation,
 reached
at $\epsilon >$1.3 GeV/fm$^3$.
Assuming that 
the 
 intermediate mass region (IMR) 
dimuon enhancement seen by NA50 is due to open
charm ($D \overline{D}$), the following observation can be made:
a) Charm is not equilibrated.
b) $J/\Psi/D \overline{D}$ suppression -unlike $J/\Psi/DY$-
appears also in S+A collisions, above $\epsilon$ $\sim$1 GeV/fm$^3$.
c) Both charm and strangeness show a discontinuity near the same $\epsilon$.
d) $J/\Psi$ could be formed mainly through $c \overline{c}$ coalescence.
e) The enhancement factors of hadrons with u,d,s,c quarks
may be connected in a simple way to the mass gain of these particles
if they are produced out of a quark gluon plasma (QGP).
We discuss these results as possible evidence for the QCD phase transition
occuring near $\epsilon \sim $1.3 GeV/fm$^3$.
}]
\section{Introduction}
\noindent
The quark-gluon plasma phase transition predicted by QCD \cite{qcd}
 may occur and manifest itself in ultrarelativistic nuclear collisions
 through discontinuities in the 
initial energy density ($\epsilon_i$) dependence of relevant observables.
A major example of a discontinuity is seen in the
 $J/\Psi/DY$ \cite{na50} discussed e.g. in \cite{satz_review,pbm}.
We investigate  for the first time the dependence of 
strangeness production, in particular of kaons, on the initial energy density
$\epsilon_i$ \cite{my_sqm2000_2,0004138}.
\noindent
The degree of equilibrium achieved in nuclear collisions has been
intensively studied comparing hadron ratios and densities to models
(see e.g. \cite{pbm,redlich,becatini,rafelski}).
We investigate here if chemical equilibrium is achieved,
examining an other aspect of equilibrium states,
namely the volume ($V$) independence of hadron densities
($\rho$).\footnote{Results 
of the NA52 experiment shown in this talk can be found
in \protect\cite{1talk,my_sqm2000_1}.}
\section{Results and discussion}
\noindent
The kaon density ($\rho_K$=(K per collision)/V) at the thermal freeze out
in nuclear reactions, investigated 
as a function of the initial energy density $\epsilon_i$ (figure 1) 
(see \cite{0004138} for calculation details),
exhibits a dramatic changeover around $\epsilon$=1.3 GeV/fm$^3$,  saturating
 for higher $\epsilon$ values, while it is falling below.
The syst. error on $\epsilon_i$ is estimated to be $\sim$ 30\%.
It is assumed that the number of
nucleons participating in the collision (N) is proportional
to the volume of the particle source at the thermal freeze out
\cite{0004138}.
The new results from Si+Au at 14.6 A GeV and p+p at 158 A GeV 
shown in figure 1, which are not included in \cite{0004138}, 
have been estimated using data  from
\cite{siau} and methods described in \cite{0004138}.
Furthermore, $\rho_K$ rises with N respectively with V below
$\epsilon$=1.3 GeV/fm$^3$ 
while it does not depend on N respectively on V above $\epsilon$=1.3 GeV/fm$^3$.
To illustrate this, two values of $V$ are noted on figure 1.
The changes of $K^{\pm}$ and $\pi^{\pm}$ with N within the Pb+Pb system, 
have been first realized in \cite{na52_centr}.
A similar behaviour as the one seen in figure 1, can be inferred for
 pions as well as for the $K/\pi$ ratio (S.K. work in progress).
\begin{figure}
\vspace*{-0.4cm}
\epsfxsize170pt
\figurebox{}{}{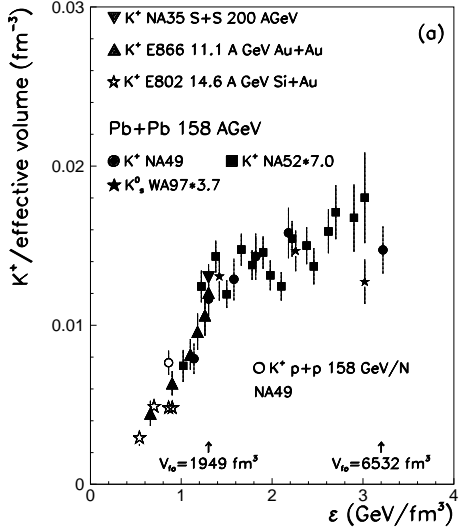}
\caption{
Initial energy density ($\epsilon$) dependence of
 the  $K^+$ multiplicity over the effective volume 
of the particle source at thermal freeze out  \protect\cite{0004138}.
}
\label{kaons}
\vspace*{-0.4cm}
\end{figure}

\vspace{0.2cm}

\noindent
The $N^{\alpha}$ exponent of hadrons with (u,d,s) quarks
above $\epsilon$=1.3 GeV/fm$^3$,
do not depend on the particle mass (figure 3).
At $\epsilon>$1.3 GeV/fm$^3$ $\alpha$ is near to one,
as expected in case of a chemically equilibrated
state, assuming N $\sim$ V.
 The deviations seen in $\phi$, $\pi^0$ and $\overline{p}$ may be 
 due to the transverse momentum acceptance.
Therefore, figure 3 supports the assumption of
 a high degree of chemical equilibrium  reached 
above $\epsilon$=1.3 GeV/fm$^3$, among hadrons with u,d,s quarks.
The
 $N^{\alpha}$ exponent of kaons
 is found to depend strongly on $\sqrt{s}$ for kaons (figure 4).
Therefore, below $\epsilon$=1.3 GeV/fm$^3$, $\rho_k$ 
(figure 1 and figure 4), 
$\rho_{\pi}$ and the $K/\pi$ ratio,
 show an increase with increasing $N$ respectively with $V$.

\begin{figure}
\vspace*{-0.4cm}
\epsfxsize170pt
\figurebox{}{}{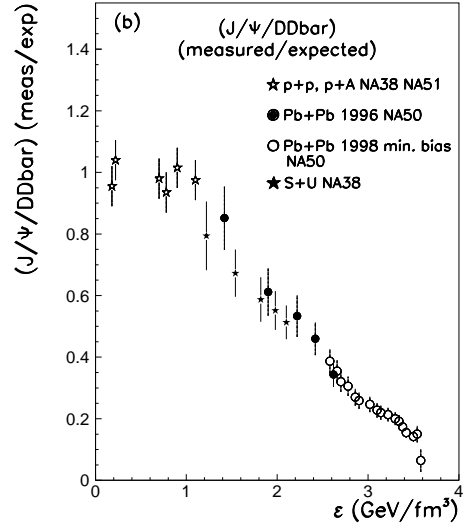}
\caption{
Initial energy density ($\epsilon$) dependence of
the $J/\Psi/D \overline{D}$ (measured/'expected') ratio
\protect\cite{0004138}.
}
\label{kaons}
\vspace*{-0.4cm}
\end{figure}

\vspace{0.2cm}

\noindent
Figures 1, 3 and 4 can be interpreted in two ways.
Firstly,
 kaons may achieve a higher degree of chemical equilibrium 
only 
for $\epsilon >$ 1.3 GeV/fm$^3$, and may not be fully equilibrated
 below \cite{0004138}.
The equilibration of strangeness is expected in a QGP 
and its observation at $\epsilon \sim $ 1.3 GeV/fm$^3$
 could therefore be a sign of a transition to QGP.
In this case, 
 it is a transition from a non equilibrated
hadron gas to an equilibrated QGP.
 It is therefore not a well defined phase
transition in the thermodynamic sense.
\vspace{0.2cm}

\noindent
Secondly,
 kaons can be in fact chemically equilibrated
also below  $\epsilon$= 1.3 GeV/fm$^3$, and the 
change respectively the constancy  of
$\rho_K$ with $V_{fo}$ and $\epsilon_i$ observed in figure 1, 
can be a result of the increase of the freeze out temperature with $\epsilon_i$
below $\epsilon$= 1.3 GeV/fm$^3$, respectively of
 the stability of $T_{fo}$ above 1.3 GeV/fm$^3$.
This dependence of $T_{fo}$ on $\epsilon_i$, namely rising until
it reaches a critical $T_c$ value and saturating above for all reactions,
would strongly support the QCD phase transition appearing 
at $\epsilon \sim$ 1.3 GeV/fm$^3$. 
\noindent
This interpretation fully agrees
 with thermal models  which suggest that particle ratios at freeze out
are compatible with thermalization
even in A+A collisions at 1 A GeV \cite{redlich}.
However the first interpretation is not in gross disagreement with
\cite{redlich}, because there  the thermal model description is
modified (introducing e.g. $\rho_k \sim V$)
 in order to describe the data at 1 A GeV.
\begin{figure}
\epsfxsize190pt
\figurebox{}{}{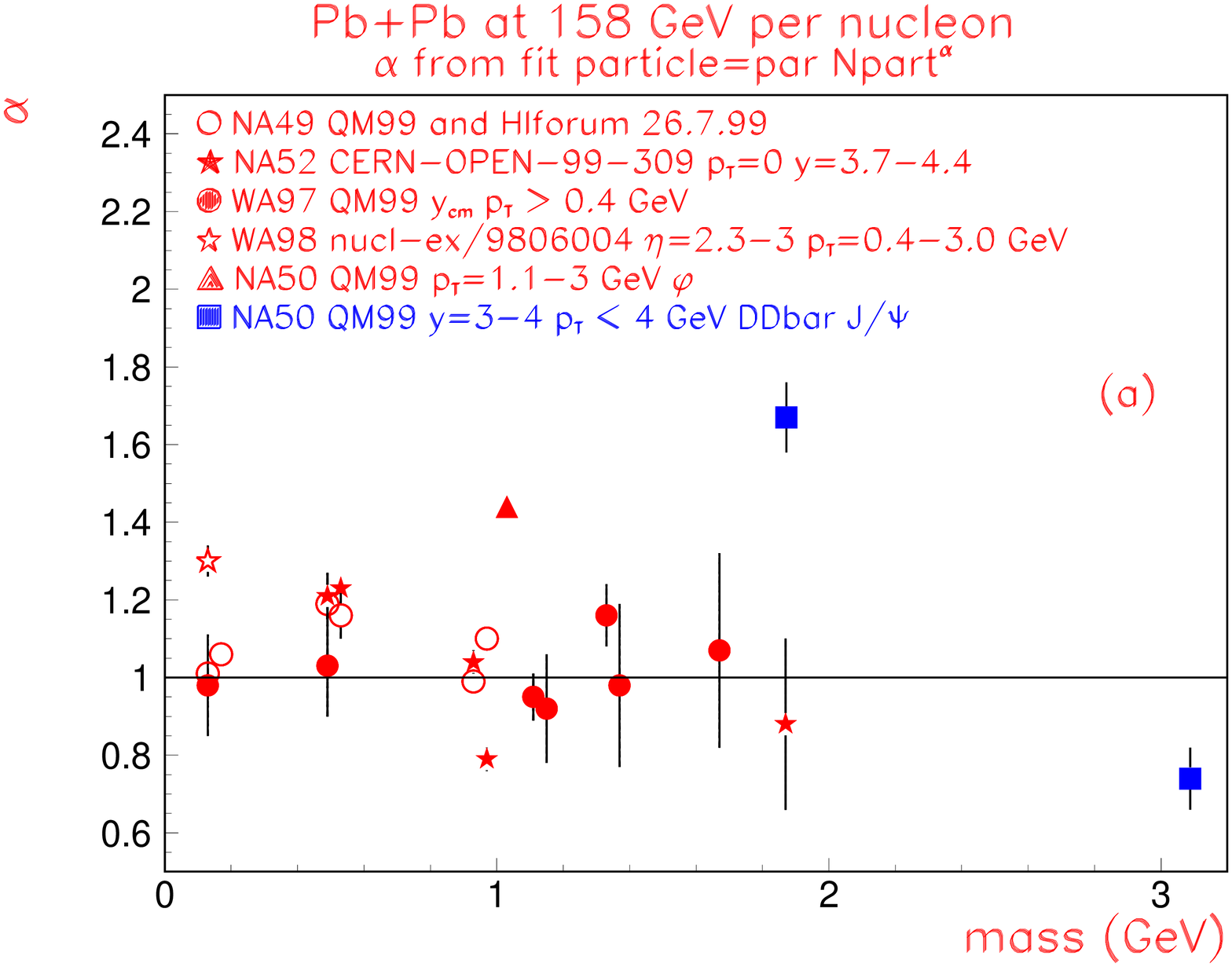}
\caption{
The parameter $\alpha$, resulting from the
$N^{\alpha}$ fit to hadron yields shown as a function of 
 the mass of the particles in the region $\epsilon>$1.3GeV/fm$^3$ at SPS.
N is the number of participating nucleons.
}
\label{kaons}
\vspace*{-0.4cm}
\end{figure}

\noindent
Furthermore, the correct interpretation can be corroborated by
further investigations discussed in the following.
The nonzero baryochemical potential ($\mu_B$), which in the reactions shown
in figure 1, happens to
change with $\epsilon_i$, makes the intepretation of figure 1 difficult.
Therefore, it appears that the
dependence of the temperature at chemical freeze out extrapolated to $\mu_b$=0,
 on $\epsilon_i$, would help to identify and prove the QCD phase transition. 
A rising and then a for ever saturating freeze out temperature
above $\epsilon$ 1.3 GeV/fm$^3$ is a strong argument
that the QCD phase transition occurs at this $\epsilon$,
and figure 1 is a direct consequence of it.

\begin{figure}
\epsfxsize190pt
\vspace*{-0.2cm}
\figurebox{}{}{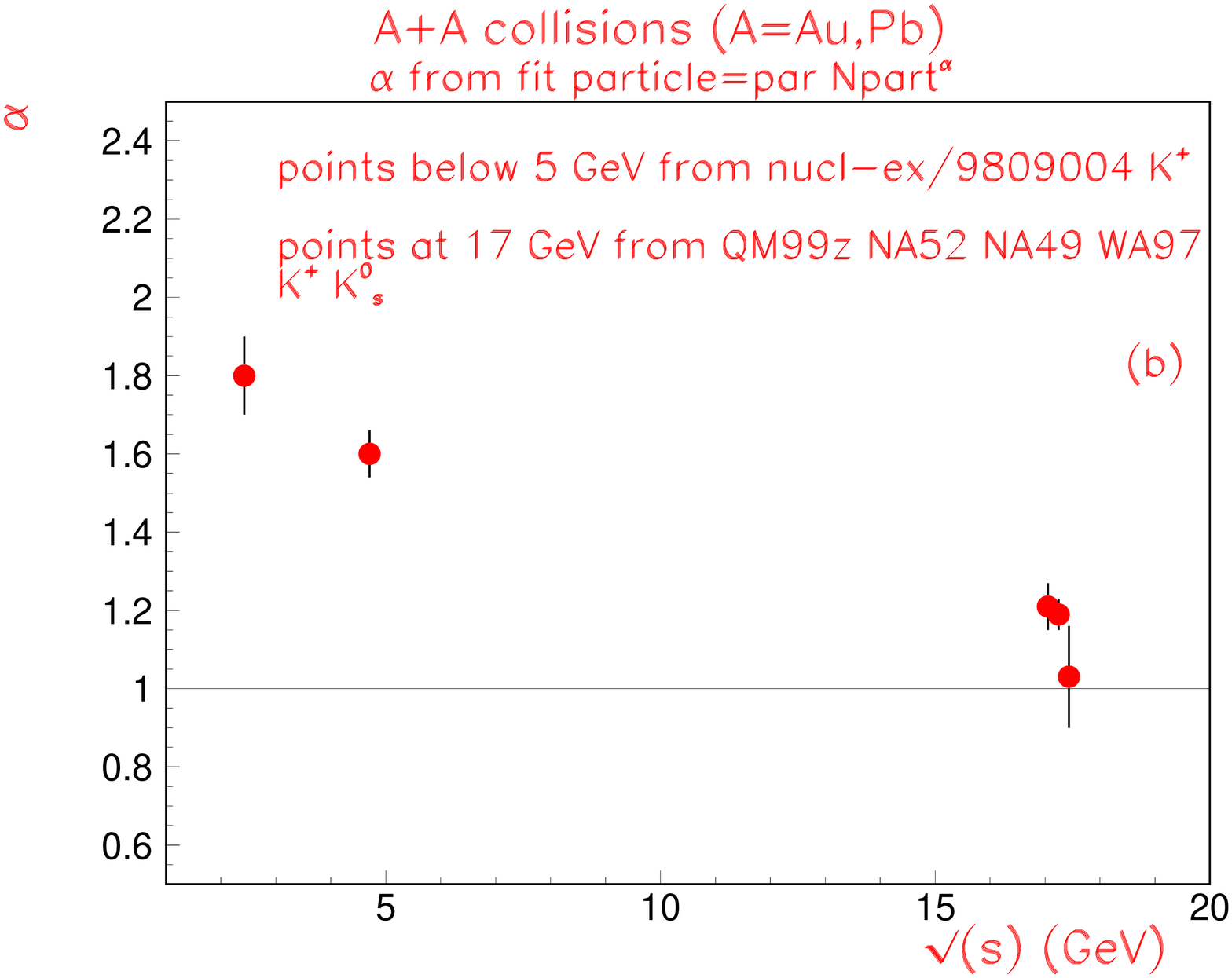}
\caption{
The parameter $\alpha$, resulting from the
$N^{\alpha}$ fit to hadron yields shown as a function of 
 $\sqrt{s}$ for kaons.
N is the number of participating nucleons.
}
\label{kaons}
\vspace*{-0.4cm}
\end{figure}
\noindent
The question if the QCD phase transition appears at the critical $\epsilon_i$
in any volume,
or if there is additionally a 
 critical initial
 volume of the particle source above which the transition takes place, can be answered
 comparing QGP signatures in systems with different volumes but the
same $\epsilon_i$.
For example comparing $p+p$, $e^+ e^-$ etc  collisions
to heavy ion collisions e.g. at the same $\epsilon$. 
This is not yet done for the signature of the $J/\Psi$ suppression 
and it has to be clarified e.g. using Tevatron data \cite{0004138}.
For the signature of strangeness enhancement 
it is suggested by figure 1 in \cite{becatini}
that there is indeed a critical initial volume,
 only above which strangeness is enhanced over $p+ \overline{p}$ 
{\it at the same $\epsilon_i$}.
This conclusion follows, if we assume that Tevatron reaches at least
$\epsilon_i$ values similar to SPS A+A collisions \cite{na35}
and if figure 1 in \cite{becatini} is not 
biased by the model calculation \cite{becatini}.
\vspace{0.2cm}

\noindent
If strangeness is indeed not equilibrated at $\epsilon < 1.3$ GeV/fm$^3$,
this may explain 
 the decrease of the double ratio ($K/\pi$)(A+A/p+p) with increasing
$\sqrt{s}$.
In particular, a larger strangeness annihilation is enforced by equilibrium
at SPS reducing the strange particle yield.
However the assumption of 
non equilibrium of $s \overline{s}$ at low $\epsilon$
 is not nessecary here, since the above observation
can be possibly traced back
 to e.g. the variation of $\mu_B(A+A) / \mu_B(p+p)$ with $\sqrt{s}$
in A+A collisions.
Furthermore,
in the context of QGP formation, it seems irrelevant to discuss
e.g.  $s \overline{s}$ enhancement in A+B over p+p collisions
 in a nonequilibrium situation.
It is the very establishment of equilibrium in the (u,d,s) sector, 
 which can reveal informations on QGP.
\\
The kaon number densities in p+p and A+B collisions
in figure 1 are similar, when compared at the same $\epsilon_i$.
See also \cite{heinz} for a discussion of universality of
 pion phase space densities.
\vspace{0.2cm}

\noindent
Our prediction for the N dependence of
hadrons at RHIC and LHC is the $N^1$ thermal limit,
as long as hadron yields are 
 dominated by low transverse momentum  particles.
Furthermore,
if the changeover of $\rho_k$ at $\epsilon=$1.3 GeV/fm$^3$ shown in
figure 1 is due to the QCD phase transition,
we predict for RHIC and LHC the same 
total strangeness (or kaon) number density and the
same freeze out temperature, -after correction for the $\mu_B$ dependence-,
as for $\epsilon=$1.3-3.0 GeV/fm$^3$.
If this change is
 however due to the onset of equilibrium in a hadronic gas, 
and the QCD phase transition takes place at 
higher $\epsilon$, it may manifest itself through a 
second changeover of hadron number densities, 
ratios and freeze out temperatures -after correction for the different
$\mu_B$-  e.g. in RHIC
above  $\epsilon $ $\sim$ 3 GeV/fm$^3$.

\vspace{0.2cm}

\noindent
Assuming that the IMR dimuon enhancement seen by NA50 is due to open
charm, the following observations can be made:
a)
open charm appears not to be equilibrated ($\alpha=1.7$) (figure 3)
 \cite{0004138}.
b) The $J/\Psi/D \overline{D}$ ratio deviates from p+p and p+A data
also in S+U collisions (figure 2), above $\epsilon$ $\sim$1 GeV/fm$^3$.
c) It therefore appears that both charm and strangeness show a discontinuity
near the same $\epsilon$$\sim$1 GeV/fm$^3$  \cite{0004138},
 similar to the critical $\epsilon_c \sim$1-2 GeV/fm$^3$ predicted by QCD
\cite{qcd,satz_review}.
d) The N dependence of the $J/\Psi/D \overline{D}$ ratio 
can be interpreted as the $J/\Psi$ being 
formed through $c, \overline{c}$ coalescence \cite{0004138}.
\vspace{0.2cm}

\noindent
e) Finally, the enhancement factors of hadrons with u,d,s,c quarks
may be connected in a simple way to the mass gain of these particles
in the quark gluon plasma (table below) \cite{skpm}.
$T_q$ are the enhancement factors of the lightest mesons with
u,d,s,c quarks ($\pi, K, D$), if they are produced out of a quark gluon 
plasma (e.g. $g + g \rightarrow s + \overline{s} $ (1)),
 as compared to their direct production from hadron interactions
away from the transition point
(e.g. $p + p \rightarrow K^+ + \Lambda +p$ (2)). 
The gain is taken proportional to $m_{particle} - m_{quarks}$, as this
expresses the different thresholds of reactions (1) and (2).
In the table below
 the predicted enhancement factors ($T_q$) of hadrons with u,d,s,c quarks 
from a QGP are compared 
to the experimentally measured ones ($E_q$), and are found to be similar.
(Definitions:
$th_q=m_0 - m_q$, $m_{u,d}$=7 MeV, $m_s$=175 MeV, $m_c$=1.25 GeV,
 $m_0=m(\pi, K, D)$,
$T_q=\sqrt{ th_q / th_{u,d}}$, $E= \frac {(A+B)} {(N+N)}$, $E_q=E/E_{u,d} $).
\begin{table}
\begin{tabular}{lllll}
$q$  & $th_q$ & $T_q$ & $E$ & $E_q$
\\
\hline
 u,d  & 133 & 1 &   $\pi/N$ $\sim$ 1.12 & 1
\\
 s    & 320 & 1.55 &  $K/N$ $\sim$ 2   & 1.79
\\
 c    & 615 & 2.15  & $D \overline{D}$ $\sim$ 3  & 2.68
\\
\end{tabular}
\end{table}
\\
\noindent
{\bf Acknowledgments}
I would like to thank Prof.~K~Pretzl and Prof.~P~Minkowski for 
stimulating discussions and 
  the Schweizerischer Nationalfonds for their support.


\begin{thebibliography}{9}
\bibitem{1talk} S.~Kabana et al (NA52 coll.), hep-ex/0010045,
  contributed to ICHEP2000, preprint BUHE-00-07.
\bibitem{my_sqm2000_1} S.~Kabana et al (NA52 coll.), hep-ex/0010053,
proceedings of the conference SQM2000 and Univ. of Bern preprint
BUHE-00-05.
\bibitem{qcd} E. Laermann, Nucl. Phys. A610 (96) 1c.
\bibitem{na50} M.Abreu et al. (NA50 coll.),CERN-EP-2000-013.
\bibitem{satz_review} H. Satz, hep-ph/0007069.
\bibitem{pbm} P. B. Munzinger et al., nucl-th/0007059.
\bibitem{redlich} J. Cleymans, H. Oeschler, K. Redlich, Phys. Lett. B485 (00) 27.
\bibitem{becatini} F. Becattini et al, hep-ph/0002267.
\bibitem{rafelski}  J. Rafelski, hep-ph/0006085.
\bibitem{0004138} S. Kabana, hep-ph-0004138.
\bibitem{my_sqm2000_2} S.~Kabana, proceedings of the conference
 SQM2000 and Univ. of Bern preprint BUHE-00-06, hep-ph/0010228.
\bibitem{siau} E802 nucl-ex/9903009,
 Nucl. Phys. A661 (99) pages: 45, 439, 448,
Phys. Lett. B332 (94) 258, 
\\
M. Tannenbaum priv. communication.
\bibitem{na52_centr}
G. Ambrosini et al. (NA52 Coll.), New J. of Phys. (1999) 1, 22 and
New J. of Phys. (1999) 1, 23,
  S. Kabana et al, (NA52 coll.) Nucl. Phys. A661 (99) 370c. 
\bibitem{na35} J. Bartke et al, (NA35 coll.) Z. Phys. C 48 (1990) 191.
\bibitem{heinz} D. Ferenc et al, Phys. Lett. B457 (99) 347.
\bibitem{skpm} S. Kabana, P. Minkowski, to be published.
\end{thebibliography}
\end{document}